\documentstyle[11pt,aaspp4,tighten]{article}

\tightenlines

\def\gsimeq
{\hbox{\raise0.5ex\hbox{$>\lower1.06ex\hbox{$\kern-1.07em{\sim}$}$}}}
\def\lsimeq
{\hbox{\raise0.5ex\hbox{$<\lower1.06ex\hbox{$\kern-1.07em{\sim}$}$}}}

\begin{document}

\title{Evidence for Nonlinear X-ray Variability from the Broad-line
Radio Galaxy 3C~390.3}

\author{Karen M. Leighly\altaffilmark{1}}\altaffiltext{1}{Current
address: Columbia Astrophysics Laboratory, 538 West 120th Street, New
York, NY 10025, USA, leighly@ulisse.phys.columbia.edu}
\affil{Cosmic Radiation Laboratory, RIKEN, Hirosawa 2--1, Wako-shi, 
Saitama 351, Japan, leighly@postman.riken.go.jp}
\author{Paul T. O'Brien}
\affil{Department of Physics \& Astronomy, University of
Leicester, University Road, Leicester, LE1~7RH, U.K., pto@star.le.ac.uk}

\slugcomment{Submitted to {\it The Astrophysical Journal Letters}}


\begin{abstract}

We present analysis of the light curve from the {\it ROSAT} HRI
monitoring observations of the broad-line radio galaxy 3C~390.3.
Observed every three days for about 9 months, this is the first well
sampled X-ray light curve on these time scales. The flares and
quiescent periods in the light curve suggest that the variability is
nonlinear, and a statistical test yields a detection with $\gsimeq 6
\sigma$ confidence.  The structure function has a steep slope $\sim
0.7$, while the periodogram is much steeper with a slope $\sim 2.6$,
with the difference partially due to a linear trend in the data.  The
non-stationary character of the light curve could be evidence that the
variability power spectrum has not turned over to low frequencies, or
it could be an essential part of the nonlinear process.  Evidence
for X-ray reprocessing suggests that the X-ray emission is not from
the compact radio jet, and the reduced variability before and after
flares suggests there cannot be two components contributing to the
X-ray short term variability. Thus, these results cannot be explained
easily by simple models for AGN variability, including shot noise
which may be associated with flares in disk-corona models or active
regions on a rotating disk, because in those models the events are
independent and the variability is therefore linear.  The character of
the variability is similar to that seen in Cygnus X-1, which has been
explained by a reservoir or self-organized criticality model.
Inherently nonlinear, this model can reproduce the reduced variability
before and after large flares and the steep PDS seen generally from
AGN. The 3C~390.3 light curve presented here is the first support for
such models to explain AGN variability on intermediate time scales
from a few days to months.

\end{abstract}


\section{Introduction}

X-ray variability has been observed in active galaxies (AGN) for two
decades, but the origin of the variability is comparatively poorly
understood. So far, the {\it EXOSAT} long observations have provided
the most constraining results on AGN variability.  On time scales $<3$
days, the power density spectrum (PDS) can be described with a power
law of slope $s > 1$ ($P(f) \propto f^{-s}$).  The index clusters
around $s=1.55$ over a large range of luminosity; further, an inverse
correlation between PDS amplitude and X-ray luminosity seems to
generally support the idea that less luminous objects have a smaller
black hole mass (Lawrence \& Papadakis \markcite{14} 1993).  On longer
time scales, the PDS must flatten to low frequencies, or the total
power would diverge.

As demonstrated by Vio et al. \markcite{29} (1992), the PDS cannot be
used to distinguish whether the process is linear or nonlinear.  This
determination is fundamental in AGN research, as it provides
information about the structure of the emitting region.  For example,
one of the most widely accepted generic models to explain AGN
variability, that the red noise power spectrum results from the
flaring of many independent active regions, is an inherently linear
model.  If nonlinearity is discovered in the light curve, it means at
least that the active regions must not be completely independent all
the time, or in the extreme case, that there is a single emission
region.

3C~390.3 is a luminous ($L_{X(2-10)}\sim 2-4 \times 10^{44} \,\rm ergs
\, s^{-1}$) nearby (z=0.057) broad-line radio galaxy located in the
North Ecliptic cap.  It is well known as the prototypical source of
broad double-peaked and variable $H\beta$ lines (e.g. Veilleux \&
Zheng \markcite{27} 1991).  3C~390.3 is variable in all wave bands,
including the optical (Barr et al.
\markcite{4} 1980) and UV (Zheng \markcite{31} 1996), and it is a
bright and variable X-ray source (see Eracleous, Halpern \& Livio
\markcite{7} 1996
for a compilation).

During 1995, 3C~390.3 was subject to a monitoring campaign from radio
through X-rays.  Using the {\it ROSAT} HRI, we obtained the first well
sampled X-ray light curve on time scales from days to months, shown in
Figure 1.  The surprising result is that there is clear evidence for
nonlinear variability in the X-ray light curve, the first such
reported for an AGN.  This result imposes new constraints on AGN
variability models.  The details of the observations, the preliminary
analysis of the {\it ROSAT} HRI data and the results from two {\it
ASCA} observations made during the monitoring period are discussed in
a companion paper Leighly et al. \markcite {15} 1997a.  The results
from the optical, UV and radio monitoring will be reported elsewhere
(Dietrich et al. \markcite{6} 1997, O'Brien et al. \markcite{22} 1997,
Leighly et al. \markcite{16} 1997b).

\begin{figure}[h]
\vbox to3.75in{\rule{0pt}{3.75in}}
\includegraphics{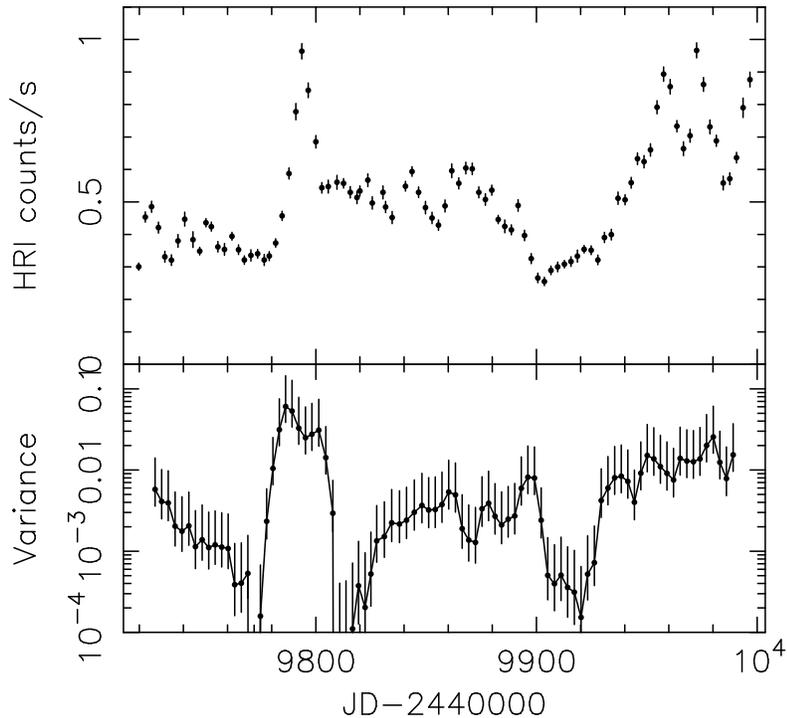}
\caption{Top: {\it ROSAT} HRI light curve from monitoring observations of
3C~390.3; Bottom: Quiescent periods in the light curve can be detected
by taking the variance over a sliding box with $N=6$.}
\end{figure}

\section{Variability Analysis}

\subsection{Evidence for Nonlinearity}

The 3C~390.3 light curve looks different than X-ray light curves from
AGN on shorter time scales (e.g. McHardy \markcite{17} 1989).  The
most noticeable features are an isolated flare at about TJD 9800, and a
series of flares starting at TJD 9960.  Qualitatively speaking, such
flares are evidence that the variability is nonlinear (e.g. Vio
\markcite{29} 1992).

Various statistical tests for nonlinearity are available.  We use the
method of Theiler et al. \markcite{26} (1992), explained fully in this
reference, which uses simulations to assess the significance of
the detection of nonlinearity in the time series.  First, a null
hypothesis is conceived, and surrogate time series are simulated which
have the properties of the null hypothesis.  Then a nonlinear
statistic is applied to the real time series and multiple realizations
of the surrogate time series. If the value of the statistic for the
real time series is significantly different than the distribution from the surrogate
data, nonlinearity is detected, and the significance of the detection
is the number of sigma difference between the real data values and
mean of the surrogate values.

The null hypothesis that we chose is that the variability is linear
but has a steep power density spectrum, as this is the here-to best
model of AGN variability (e.g. Lawrence \& Papadakis \markcite{14}
1993).  Surrogate data are generated by randomizing the Fourier phases
but retaining the amplitudes of the discrete power spectrum obtained
from the real data.  Thus the surrogate data has the same power
spectrum, circular autocorrelation, mean and variance as the real
data, but has been linearized by the phase randomization.  This
takes advantage of the fact that the power spectrum measures only the
linear properties, or the first two moments of the process (e.g. Vio
et al. \markcite{29} 1992); thus linear and nonlinear time series can
have the same power spectra.

Ideally, any nonlinear statistic can be used; however, some are more
sensitive than others for our data.  For example, the slope of the
correlation integral is a useful statistic, since the asymptotic value
is the dimension of the strange attractor (Grassberger \& Procaccia
\markcite{10} 1983).  However, after ignoring the first point and last
four points of the light curve to reduce spurious high frequency
components, there are only 88 points in our time series, and the slope
of the correlation integral is noisy and not well defined.  Instead,
we use the value of the correlation integral $C(r)$ at a specific
value of $r$ (Theiler et al. \markcite{26} 1992).  The correlation
integral counts the number of pairs of vectors with difference vector
magnitude less than $r$; this is generally larger for the optimal $r_{in}$
if the process is nonlinear.  For small $r$, the correlation becomes
zero for larger dimensions, while for large $r$, the majority of pairs
of vectors are counted, and since the observed and surrogate data are
both normalized, the correlation integrals become indistinguishable.
Figure~2, upper panel, shows the value of the correlation integral at
$r_{in}=0.5$ versus embedding dimension for the real data and for 10
realizations of the surrogate data. The lower panel of Figure 2 shows
that the detection significance of the nonlinearity is $\gsimeq
6\sigma$ computed from 100 realizations of the surrogate data.

The detection of nonlinearity is corroborated by the presence of
quiescent periods associated with intermittency in the light curve.
Examination of the light curve shows that the variability before and
after the flares seems to be reduced.  Since the signal-to-noise of
this time series is very high ($\sim 30$), this can be quantified by
computing the true variance over a sliding box (e.g. Isliker \& Benz
\markcite{13} 1994) as shown in the lower panel of Figure~1.

\begin{figure}[h]
\vbox to3.75in{\rule{0pt}{3.75in}}
\includegraphics{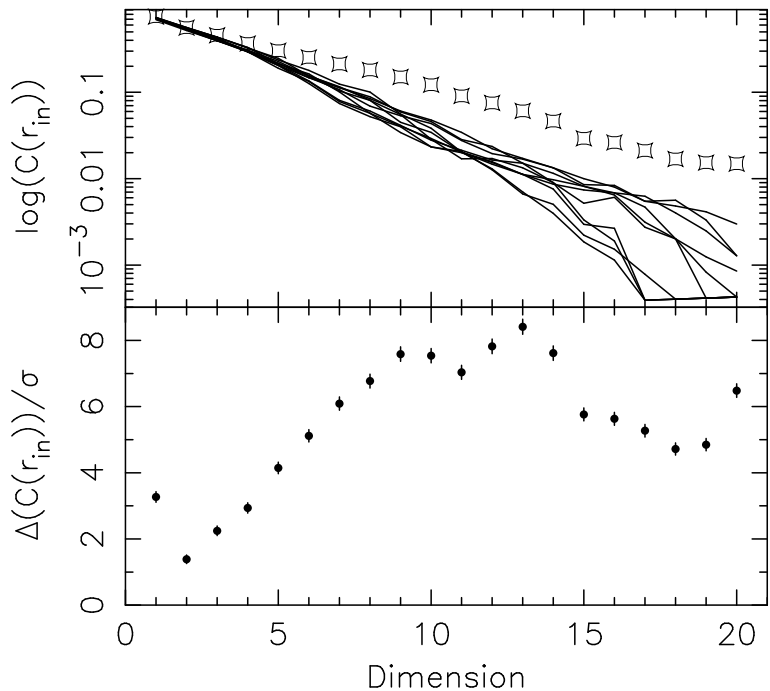}
\caption{Detection of nonlinearity in the 3C~390.3 data, using the
method of Theiler et al. 1992.  Top: Squares show value of the
correlation integral at $r_{in}=0.5$ for the real data, while lines
connect points from the simulated data; Bottom: Number of sigma
difference between values from real data and mean of surrogate data.}
\end{figure}

\subsection{The Variability Power Spectrum}  

Since the sampling is effectively even, the periodogram can be
computed directly using the discrete Fourier transform.  However,
estimating power density spectrum (PDS) parameters is complicated by
several problems.  The time period spanned is short compared with the
time scale of variability, so the value of average quantities can be
estimated only poorly.  Also, this light curve is clearly not
stationary (see e.g.  Papadakis \& Lawrence
\markcite{24} 1995), so strictly speaking, the power density spectrum
cannot be estimated from the periodogram.  Furthermore, extensive
simulations made by Papadakis \& Lawrence \markcite{24} (1995)
demonstrate with few numbers of points, problems including red-noise
leak of power toward high frequencies and aliasing toward low
frequencies skew the periodogram severely.

The periodogram computed using the method of Papadakis \& Lawrence
\markcite{25} (1993) is shown in the upper panel of Figure 3.  A power
law fit to the binned periodogram excluding the first and last points
yields $s=2.7$. Similar results are obtained if a spectral window is
applied to the light curve ($s=3.0$).  This slope is much steeper than
those found at higher frequencies using {\it EXOSAT}.

The structure function (SF) slope $d$ is related to the slope of the
power density spectrum by $s=1+d$ (e.g. Hughes, Aller \& Aller
\markcite{13} 1992) again under the condition that the time series is
stationary.  The SF is shown in the lower panel of Figure 3.  For lags
between 6 and 64 days, the SF has a logarithmic slope of $d=0.7$,
corresponding to a PDS index of $s=1.7$, close to the average high
frequency slope of 1.55 found using {\it EXOSAT} data (Lawrence \& Papadakis
1993).  Simulating 1000 light curves with PDS index $s=1.7$ and
similar length, and fitting over the same interval gives mean $d=0.58$
with standard deviation 0.27.  Part of the difference between the
periodogram and the structure function can be traced to the
nonstationarity.  The structure function slope ignores low
frequencies, but the periodogram will be steepened by an overall trend
in the data.

While information regarding the PDS is only poorly estimated using
these data, note that the nonstationarity of the light curve could be
evidence that the PDS has not turned over toward low frequencies
indicating that the break in the power spectrum occurs on longer time
scales than about 4 months for this comparatively luminous object.
Alternatively, as nonlinear processes need not be intrinsically
stationary, the power spectrum actually could be changing with time.

\begin{figure}[h]
\vbox to5.5in{\rule{0pt}{5.5in}}
\includegraphics{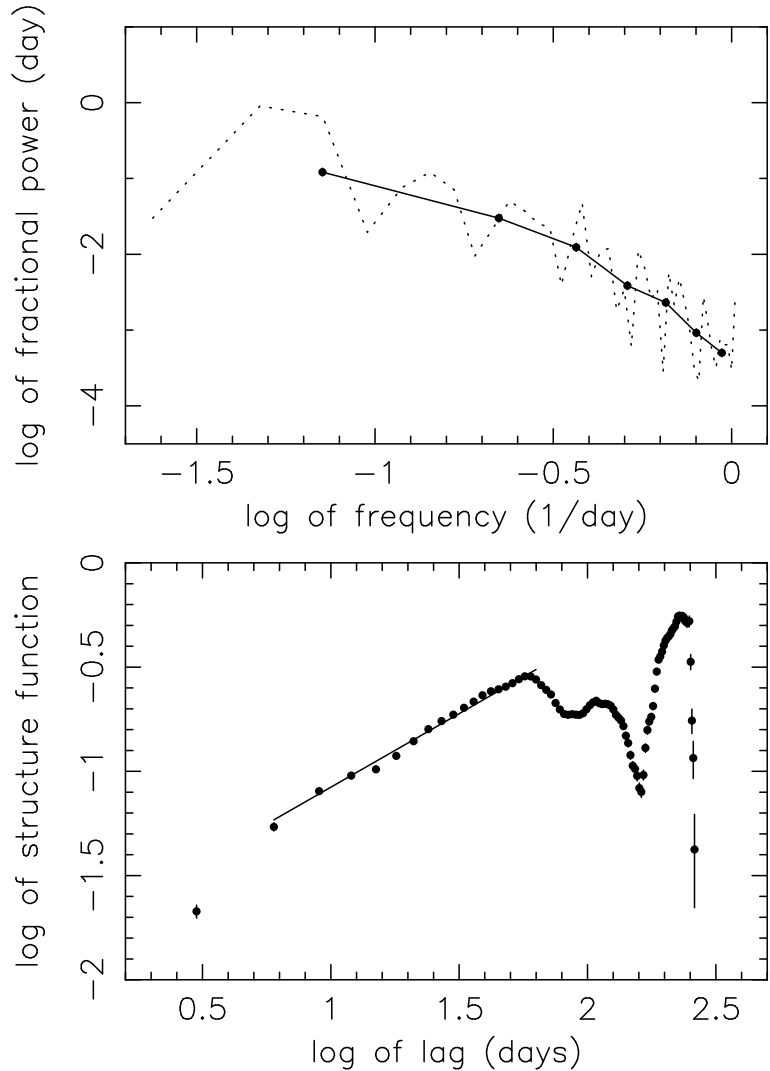}
\caption{Top: the periodogram; Bottom: the structure function.} 
\end{figure}

\section{Discussion}

\subsection{An X-ray jet in 3C~390.3?}

3C~390.3 has a compact radio jet (Alef et al. \markcite{1} 1996) and
it is possible that the X-ray emission comes from the jet.  If so,
nonlinear X-ray variability may be expected since essentially a single
emission region would be present.  Furthermore, nonlinear variability
has been found in the optical light curves from the OVV blazar 3C~345
(Vio et al. \markcite{18} 1991).

However, there is spectral evidence that in 3C~390.3 the X-ray emission is
predominantly isotropic.  A broad iron $K\alpha$ line, similar to
those found in radio-quiet Seyfert~1 galaxies (e.g. Fabian et al.
\markcite{8} 1995) has been detected in the X-ray spectrum
(Eracleous, Halpern \& Livio \markcite{11} 1996).  These broad iron
lines are strong evidence for illumination of a relativistic accretion
disk by the X-ray source (Tanaka et al.
\markcite{25} 1995) and they are never seen in the X-ray spectra from
blazars.  Also, a weak X-ray reflection component, generally thought
to be the signature of illumination of optically thick material by
X-rays (e.g.  George \& Fabian \markcite{9} 1991) has been observed in
this object (Wozniak et al. \markcite{30} 1997).

It is also possible that there are two highly variable X-ray emission
components, one approximately isotropic with linear variability and
the other from the jet, contributing superimposed flares.  However,
the reduction of variability in the quiescent periods before and after
flares argues against this possibility.

\subsection{Current X-ray Variability Models of AGN}
 
Disk-corona models, in which the X-ray emission originates from
comptonization of soft UV thermal photons in a corona of hot electrons
lying above the disk, can successfully explain the UV--$\gamma$-ray
broad band emission of radio-quiet AGN (e.g. Haardt
\& Maraschi \markcite{11} 1993). To obtain the observed photon indices,
$\gamma$-ray cutoffs and UV to X-ray ratios, the corona should not
cover the disk completely, but rather be localized to a number of
active blobs (Haardt, Maraschi \& Ghisellini \markcite{12} 1994).  The X-ray
variability would be stochastic if it is associated with variations in
the number and luminosity of the independent active blobs.  This
situation can be modeled by shot noise, the sum of randomly occurring
independent flares with perhaps variable exponential decay times,
which can also explain the $1/f^s$ variability observed on time scales
less than 1 day (e.g.  Lawrence \& Papadakis \markcite{14} 1993).
However, shot noise is a linear model, being simply a sum of
independent events, and thus it cannot reproduce the light curve
presented here.  Similarly, a $1/f^s$ PDS can also be produced by the
differential rotation of hot spots on a disk where the observed
emission has been modified by gravitational
effects (e.g. Bao \& Abramowicz \markcite{3} 1996), but again if the
emission from each blob is independent, the variability should be
linear.  

\subsection{Analogy with Cygnus X-1 and Reservoir Models}

The Galactic black hole candidate Cyg X-1 in the low state has many
similarities with radio-quiet AGN, including a $1/f^s$ power density
spectrum.  Recently, Negoro and collaborators have attempted to
understand the variability of this object by identifying the largest
flares in the light curves and stacking them to determine common
properties.  They find that the probability of flaring is
significantly reduced before and after large flares (Negoro et al.
\markcite{19} 1995).  The quiescence we observe before and after
flares in the 3C~390.3 light curve seems to be a similar phenomenon.

A self-organized criticality (SOC) disk model was developed to explain
these results (Mineshige, Ouchi, \& Nishimori \markcite{18} 1994).
This model uses the concepts of self-organized criticality proposed by
Bak et al. \markcite{2} (1988) to explain $1/f$ variability observed
in a wide variety of physical systems.  The SOC accretion disk is
comprised of numerous reservoirs.  When a critical density is reached
in a reservoir, an unspecified instability is activated causing an
avalanche of accretion and emptying of the reservoir.  Adjacent
reservoirs are coupled, so that triggering instability in one
reservoir may result in instability in few or many adjacent
reservoirs, resulting in a small or large flare. It is this coupling
which provides the essential nonlinearity in the model. In this model,
small flares occur randomly, since they originate from just a few
reservoirs.  Quiescent periods after large flares naturally occur
while the large number of reservoirs emptied fill again.  Quiescent
periods before large flares happen because few or no small flares have
occurred to release the accumulated potential energy.  The predicted
power density spectra are steep ($\propto 1/f^{1.6-1.8}$; Mineshige,
Takekuchi \& Nishimori \markcite{19} 1994) similar to the estimate
from the structure function found here, but also to the periodogram
estimates from {\it EXOSAT} light curves (Lawrence \& Papadakis 1993).
A reservoir model has been discussed in the context of AGN by Begelman
\& De~Kool \markcite{5} (1990).

The analogy with 3C~390.3 may be quite direct.  We note that the FWHM
of the flare at TJD 9800 is roughly 12 days, while the typical FWHM of
the shots in Cygnus X-1 is 0.2s (Negoro, Miyamoto \& Kitamoto,
\markcite{20} 1994).  The ratio of these timescales is $\sim 5 \times
10^6$.  The mass of the black hole in Cygnus X-1 is thought to be
about $10\rm M_\odot$.  If the flare time scale is directly
proportional to the black hole mass, a mass of $\sim 5 \times 10^7
\,\rm M_\odot$ would be implied, which is reasonable for this luminous
object.

\acknowledgements

The authors thank the referee, Roberto Vio, for many useful comments
especially regarding estimation of the PDS slope.  K.M.L. thanks Chris
Done, Tahir Yaqoob and Greg Madejski for useful discussions, and
Randall A.  LaViolette for pointing toward Theiler's method.  K.M.L
gratefully acknowledges support by NASA grant NAG~5-2637 (ROSAT) and
through a Science and Technology Agency postdoctoral fellowship.

\newpage

\clearpage

\end{document}